\documentclass[twocolumn]{article}
\usepackage{parskip}
\usepackage{amssymb}
\usepackage{amsmath}
\usepackage{fullpage}
\usepackage{authblk}
\usepackage{blindtext}
\usepackage{siunitx}
\usepackage{float}
\usepackage{caption}
\usepackage[superscript,nomove]{cite}
\usepackage{hyperref}
\usepackage[normalem]{ulem}
\usepackage{graphicx}
\usepackage{cancel}
\usepackage{algorithm}
\usepackage[noend]{algpseudocode}
\usepackage{tikz}

\makeatletter \renewcommand{\@citess}[1]{\textsuperscript{[#1]}} \makeatother

\makeatletter
\def\@firstoftwo@second#1#2{%
  \def\temp##1.##2\@nil{##2}%
   \temp#1\@nil}
\newcommand\sref[1]{%
   (A.\expandafter\@setref\csname r@#1\endcsname\@firstoftwo@second{#1})%
}

\makeatother

\title{Kinetic Monte Carlo prediction of the morphology of pentaerythritol tetranitrate}
\author[1,2]{\small Jacob Jeffries \thanks{jwjeffr@g.clemson.edu}}
\author[2]{\small Himanshu Singh}
\author[2]{\small Romain Perriot}
\author[2]{\small Christian Negre}
\author[2]{\small Antonio Redondo}
\author[1,3]{Enrique Martinez \thanks{enrique@clemson.edu}}
\affil[1]{Department of Materials Science and Engineering, Clemson University, Clemson, SC 29634, USA}
\affil[2]{Theoretical Division, Los Alamos National Laboratory, Los Alamos, NM 87545, USA}
\affil[3]{Department of Mechanical Engineering, Clemson University, Clemson, SC 29634, USA}
\date{\small \today}

\renewenvironment{abstract}
 {\quotation\small\noindent\rule{\linewidth}{.5pt}\par\smallskip
  {\centering\bfseries\abstractname\par}\medskip}
 {\par\noindent\rule{\linewidth}{.5pt}\endquotation}

\begin{document}

\twocolumn[
  \begin{@twocolumnfalse}
  \maketitle
    \begin{abstract}
        In this work, we develop an atomistic, graph-based kinetic Monte Carlo (KMC) simulation routine to predict crystal morphology. Within this routine, we encode the state of the supercell in a binary occupation vector and the topology of the supercell in a simple nearest-neighbor graph. From this encoding, we efficiently compute the interaction energy of the system as a quadratic form of the binary occupation vector, representing pairwise interactions. This encoding, coupled with a simple diffusion model for adsorption, is then used to model evaporation and adsorption dynamics at solid–liquid interfaces. The resulting intermolecular interaction-breaking energies are incorporated into a kinetic model to predict crystal morphology, which is implemented in the open-source Python package \textit{Crystal Growth Kinetic Monte Carlo} (cgkmc). We then apply this routine to pentaerythritol tetranitrate (PETN), an important energetic material, showing results in excellent agreement with the attachment energy model.
    \end{abstract}
  \vspace{0.5cm}
  \end{@twocolumnfalse}
]

\section{Introduction}

The sensitivity and performance of crystals of energetic materials are strongly dependent on the morphology of said crystals.\cite{10.1063/1.5005997, LEEPERRY2018171} For instance, close-to-spherical and smooth morphologies of pentaerythritol tetranitrate (PETN)\cite{TISDALE2022152} and cyclotrimethylene trinitramine (RDX)\cite{doi:10.1021/cg049965a, vandersteen1989} show lower shock sensitivity and are less prone to accidental detonation, while needle-like crystals show increased sensitivity,\cite{C6RA26920F,Davis2025} which can cause safety concerns. Further, crystal morphology significantly impacts performance, since bulk packing density is a strong function of particle morphology, and higher packing densities increase detonation velocities and pressures.\cite{KESHAVARZ200531, Mathieu2012} It is therefore critical to be able to predict and control crystal morphologies in order to improve the safety and performance of energetic materials—a concern that also applies to pharmaceutical compounds, which can have their properties markedly altered by the shape and size of the crystals.\cite{COOMBES20021652,Cuppen2004,Mirza2009}

Crystal morphology is often predicted using the Wulff construction method, where crystallographic planes are cut at distances proportional to geometric or energetic quantities such as the inverse of interplanar spacings (BFDH model),\cite{donnay1937new} surface energies (SE model),\cite{wulff1901xxv} or attachment energies (AE model).\cite{HARTMAN1980145} However, all these models deviate from experimental morphologies: the BFDH model highlights prominent planes but neglects energetics, whereas the SE and AE models include energetics and yield better estimates, yet still fail to capture true morphologies.\cite{Davis2025, doi:10.1021/acs.cgd.3c01487}

The SE model ensures that, at a fixed volume, the crystal shape minimizes total surface energy, yielding the equilibrium morphology.\cite{HOFFMAN1972368} However, it does not account for growth kinetics, which can kinetically trap crystal forms.\cite{CARTER19954309} The AE model is a widely used alternative to the SE model. It considers the energy released when adsorbing molecules attach to a plane,\cite{HARTMAN1980145} and has been shown to predict non-equilibrium morphologies.\cite{winn2000modeling, annurev:/content/journals/10.1146/annurev-matsci-071312-121623} However, like the SE model, it is highly sensitive to the chosen set of planes. This limitation is especially severe for low-symmetry lattices with many non-equivalent planes, such as $\beta$-1,3,5,7-tetranitro-1,3,5,7-tetrazocane ($\beta$-HMX).\cite{doi:10.1021/la802494b} In addition, small crystals are strongly influenced by edge and corner effects, which no Wulff-based method captures.\cite{WANG1984609}

Furthermore, the SE and AE models predict steady-state morphologies without access to intermediate growth dynamics, meaning neither model can compute kinetic parameters such as growth rates. To address this, Zepeda-Ruiz et al. used a Monte Carlo scheme to simulate facet evolution in PETN, in agreement with experiment.\cite{ZEPEDARUIZ2006461} However, this scheme is based on the Metropolis–Hastings algorithm, which samples state space according to the Boltzmann distribution without accounting for kinetics,\cite{10.1063/1.1699114} and thus cannot capture true crystal growth kinetics.

In this work, we introduce a kinetic Monte Carlo (KMC)-based simulation technique to predict crystal morphology using only intermolecular interaction parameters. We then implement this technique in an open-source Python package, \verb|cgkmc|.\cite{cgkmc} Lastly, we apply the technique to simulate the growth of PETN, showing excellent agreement with predictions from the AE model and experimental observations.

\section{The kMC model}

In our routine, we encode our system on a ``colored'', undirected graph $G = (V, \mathcal{Q})$, where $V$ is a set of nodes and $\mathcal{Q}$ is a set of edges. Each node $i\in V$ represents a supercell lattice site, and is equipped with a dynamical color $x_i$, representing the occupation number of site $i$:

\begin{equation}
    x_i = \begin{cases}
        1 & \text{node $i$ is occupied by a molecule} \\
        0 & \text{else}
    \end{cases}
\end{equation}

and each pair of neighboring nodes $(i, j)\in \mathcal{Q}$ is equipped with a static weight $Q_{ij} > 0$, representing the interaction energy between nodes $i$ and $j$:

\begin{equation}\label{eq:interaction-matrix}
    Q_{ij} = \begin{cases}
        \varepsilon_1 & 0 < d_\text{min}(\mathbf{r}_i, \mathbf{r}_j) \leq \delta_1 \\
        \varepsilon_2 & \delta_1 < d_\text{min}(\mathbf{r}_i, \mathbf{r}_j) \leq \delta_2 \\
        \vdots \\
        \varepsilon_m & \delta_{m-1} < d_\text{min}(\mathbf{r}_i, \mathbf{r}_j) \leq\delta_m \\
        0 & d_\text{min}(\mathbf{r}_i, \mathbf{r}_j) > \delta_m
    \end{cases}
\end{equation}

where $m$ is the shell beyond which the interaction is negligible and taken as $0$, $\delta_\ell$ is the upper bound distance on the $\ell$'th neighbor shell, and $d_\text{min}(\mathbf{r}_i, \mathbf{r}_j)$ is the minimum-image distance between nodes $i$ and $j$. Note that $Q_{ij} = Q_{ji}$, i.e. $\mathbf{Q}$ is symmetric. 

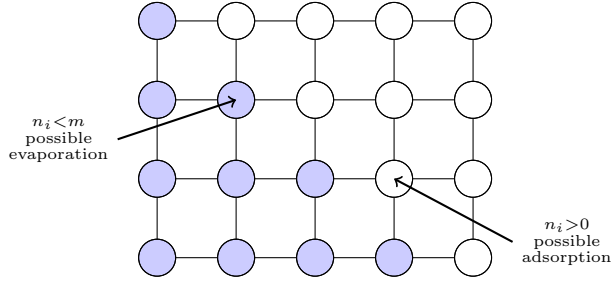
\begin{figure}[H]
    \centering
    \begin{tikzpicture}[scale=0.35]
    
        \def\n{3}
        \def\m{4}
        \def\spacing{3}
        
        \foreach \x in {0,1,...,\m} {
            \foreach \y in {0,1,...,\n} {
    
                \ifnum\x < \m
                    \draw[black] (\x*\spacing,\y*\spacing) -- (\x*\spacing+\spacing,\y*\spacing);
                \fi
                \ifnum\y < \n
                    \draw[black] (\x*\spacing,\y*\spacing) -- (\x*\spacing,\y*\spacing+\spacing);
                \fi
            
                \pgfmathtruncatemacro{\condition}{\x + \y < 4 ? 1 : 0}
                \ifnum\condition=1
                    \fill[blue!20] (\x*\spacing,\y*\spacing) circle (20pt);
                \else
                    \fill[white] (\x*\spacing,\y*\spacing) circle (20pt);
                \fi
        
                \draw[black] (\x*\spacing,\y*\spacing) circle (20pt);
                \pgfmathsetmacro\labelnum{\x*\n + \y}
                \draw[black] (\x*\spacing,\y*\spacing) circle (20pt);
                \pgfmathsetmacro\labelnum{\x*\n + \y}
            }
        }

        \coordinate (solid-pt) at (1*\spacing, 2*\spacing);
        \coordinate (solid-lbl) at (-0.5*\spacing, 1.5*\spacing);

        \node[anchor=east] at (solid-lbl) {\footnotesize $\substack{n_i < m \\ \text{possible} \\ \text{evaporation}}$};
        \draw[->, thick] (solid-lbl) -- (solid-pt);

        \coordinate (liquid-pt) at (3*\spacing, 1*\spacing);
        \coordinate (liquid-lbl) at (4.5*\spacing, 0.2*\spacing);

        \node[anchor=west] at (liquid-lbl) {\footnotesize $\substack{n_i > 0 \\ \text{possible} \\ \text{adsorption}}$};
        \draw[->, thick] (liquid-lbl) -- (liquid-pt);

    \end{tikzpicture}
    \caption{Example colored graph. Blue nodes represent sites $i$ with occupation $x_i = 1$, and white nodes represent sites $i$ with occupation $x_i = 0$. Coordination numbers $n_i$ determine the solid-liquid interface, i.e. possible candidate events.}
    \label{fig:colored-graph}
\end{figure}

Then, for a given state vector $\mathbf{x}\in\{0,1\}^n$, where $n$ is the number of lattice sites in the supercell, the total interaction energy is a quadratic form of the system's occupation numbers:

\begin{equation}
    E(\mathbf{x}) = \sum_{\substack{(i, j)\in\mathcal{Q} \\ x_i = x_j = 1}}\sum_{i < j}Q_{ij} = \frac{1}{2}\mathbf{x}^\intercal\mathbf{Q}\mathbf{x}
\end{equation}

which is notably the objective function of a quadratic binary programming optimization problem\cite{Kochenberger2014}. Additionally, we can compute the corresponding adjacency matrix of $G$:

\begin{equation}
    A_{ij} = \begin{cases}
        1 & \text{nodes $i$ and $j$ interact} \\
        0 & \text{else}
    \end{cases}
\end{equation}

where $\text{nodes $i$ and $j$ interact}$ if and only if $Q_{ij} \neq 0$. From the adjacency matrix, we can compute the coordination number at each node $i$:

\begin{equation}
    n_i = \sum_j A_{ij}x_j
\end{equation}

or equivalently $\mathbf{n} = \mathbf{A}\mathbf{x}$. This lets us easily calculate the lattice sites on the surface. For a bulk coordination number $m$, the solid and liquid surface sites satisfy $0 < n_i < m$. For example, for a two-dimensional simple cubic lattice (Fig. \ref{fig:colored-graph}) with first-nearest neighbor interactions, the solid sites on the surface are the occupied sites such that $n_i < 4$, while the liquid sites on the surfaces are the unoccupied sites such that $n_i > 0$.

From the coordination numbers $n_i$, we calculate the sites at which we can evaporate and the sites at which we can adsorb. In essence, if $n_i < m$ and $x_i = 1$, then site $i$ is a candidate for evaporation. Similarly, if $n_i > 0$ and $x_i = 0$, then site $i$ is a candidate for adsorption.

Further, assuming that the crystal is spherical with radius $R$ and surface density $\sigma$, and that the diffusivity and solubility limit of the solute in the solvent are respectively $D$ and $n_\infty$, the adsorption rate at the solvent interfacial sites is:

\begin{equation}\label{eq:ads-rate}
    R_i^{(\text{dep})} = \frac{Dn_\infty}{\sigma R}
\end{equation}

which is derived in Sec.~\ref{sec:ads-rate}. At the solid interfacial sites, the evaporation rates are given by harmonic transition state theory\cite{htst}:

\begin{equation}\label{eq:evap-rate}
    \begin{aligned}
        R_i^{(\text{evap})} &= \nu_t \exp\left(-\beta\Delta E_i^{(\text{evap})}\right)\\
        &= \nu_t \exp\left(\beta \mathbf{Q}_i\cdot\mathbf{x}\right)
    \end{aligned}
\end{equation}

where $\Delta E_i^{(\text{evap})}$ is the energy barrier of evaporation, which we take to be the energy loss from the intermolecular interactions broken at site $i$, $\beta = 1/k_BT$, $k_B$ is the Boltzmann constant, $T$ is temperature, and $\nu_t$ is the rate prefactor which is dynamically updated to ensure that the simulation reaches equilibrium at a user-specified crystal size. Here, $\mathbf{Q}_i\cdot\mathbf{x} = -\Delta E_i^{(\text{evap})}$, where $\mathbf{Q}_i$ is the $i$'th column of $\mathbf{Q}$. A detailed derivation of this shortcut is shown in Sec.~\ref{sec:evap-barrier}.

To dynamically update $\nu_t$, we set the average evaporation rates and average adsorption rates equal at the desired crystal radius $R_*$:

\begin{equation}
    \frac{Dn_\infty}{\sigma R_*} = \nu_t \left\langle \exp(\beta\mathbf{Q}_i\cdot\mathbf{x})\right\rangle
\end{equation}

which yields the following update rule, assuming that the crystal is spherical with desired size $N_*$ and number density $\rho$:

\begin{equation}\label{eq:evaporation-prefactor}
    \nu_t = \kappa\left\langle \exp(\beta \mathbf{Q}_i\cdot\mathbf{x})\right\rangle^{-1}
\end{equation}

where $\kappa$ is defined to be a collection of constants for brevity:

\begin{equation}\label{eq:kappa}
    \kappa = \sqrt[3]{\frac{4}{3}\pi\rho}\frac{Dn_\infty} {\sigma N_*^{1/3}}
\end{equation}

and $N_*$ and $R_*$ are related by:

\begin{equation}
    N_* = \rho\cdot\frac{4}{3}\pi R_*^3
\end{equation}

where $\frac{4}{3}\pi R_*^3$ is the total volume of the crystal. From an initial seed $\mathbf{x}$, the interaction matrix $\mathbf{Q}$, and the corresponding rates $R_i^{(\text{dep})}$ and $R_i^{(\text{evap})}$, we can perform the residence-time algorithm\cite{BORTZ197510} to time-evolve $\mathbf{x}$ to simulate adsorption and evaporation (Alg. \ref{alg:kmc-simplified}).

\begin{algorithm}[H]
\begin{algorithmic}[1]
\Procedure{Grow}{$\mathbf{Q}, (\mathbf{r}_1, \cdots, \mathbf{r}_n)$}
    \State $\mathbf{x} \leftarrow \text{InitialConfiguration}(\mathbf{r}_1, \cdots, \mathbf{r}_n)$
    \State $t \leftarrow 0$ \Comment{Physical time}
    \State $i \leftarrow 0$ \Comment{RTA iteration}

    \While{$i < I$}
        \State $\mathbf{n}\leftarrow \mathbf{A}\mathbf{x}$ \Comment{coord nums}
        \State $\mathbf{z}^{(1)} \leftarrow \mathbf{x}\odot\left(m\mathbf{1} -\mathbf{n}\right)$ \Comment{sol. at interface}
        \State $\mathbf{z}^{(0)} \leftarrow (\mathbf{1} - \mathbf{x})\odot\mathbf{n}$ \Comment{liq. at interface}
        \State $s^{(\text{evap})} \leftarrow \{ k \; | \; z_k^{(1)} > 0 \}$ \Comment{evap. events}
        \State $s^{(\text{ads})} \leftarrow \{ k \; | \; z_k^{(0)} > 0 \}$ \Comment{ads. events}
        \State $\nu_t \leftarrow \kappa \left\langle \exp(\beta\mathbf{Q}_k\cdot\mathbf{x})\right\rangle^{-1}_{k\in s^{(\text{evap})}}$
        \State $R_k^{(\text{evap})}\leftarrow \nu_t \exp(\beta\mathbf{Q}_k\cdot\mathbf{x})$ for all $k\in s^{(\text{evap})}$
        \State $R\leftarrow \left(\frac{3\|\mathbf{x}\|_1}{4\pi\rho}\right)^{1/3}$ \Comment{effective radius}
        \State $R_k^{(\text{ads})}\leftarrow \frac{Dn_\infty}{\sigma R}$ for all $k\in s^{(\text{ads})}$
        \State Concatenate rates $\mathbf{R}\leftarrow \left(\mathbf{R}^{(\text{evap})}, \mathbf{R}^{(\text{ads})}\right)$
        \State Select event $k$ with probabilities $\mathbf{R} / \sum_j R_j$
        \State $x_k \leftarrow 1 - x_k$ \Comment{evaporate or adsorb}
        \State $i \leftarrow i + 1$
        \State Draw $\Delta t$ from $\text{Exp}\left(\sum_j R_j\right)$
        \State $t\leftarrow t + \Delta t$
    \EndWhile
\EndProcedure
\end{algorithmic}
\caption{KMC routine}\label{alg:kmc-simplified}
\end{algorithm}

In Algo. \ref{alg:kmc-simplified}, $(\mathbf{r}_1, \cdots, \mathbf{r}_n)$ denotes the sequence of lattice positions, $\text{InitialConfiguration}(\mathbf{r}_1, \cdots, \mathbf{r}_n)$ is an arbitrarily chosen seed for the crystal growth, $t$ is the physical time within the routine, $i$ denotes the iteration variable in the KMC routine, $I$ is the number of iterations, $\odot$ denotes the Hadamard product, $m$ is the coordination number of the bulk solid phase, $\mathbf{z}^{(1)}$ and $\mathbf{z}^{(0)}$ are respectively Boolean labels for the solid and liquid sites at the interface, $s^{(\text{evap})}$ and $s^{(\text{ads})}$ are respectively the set of possible evaporation and adsorption sites, $R_k^{(\text{evap})}$ and $R_k^{(\text{ads})}$ respectively denote the rate of an evaporation and adsorption event at site $k$, $R$ denotes the deffetive radius of the crystal, and $\mathbf{R}$ denotes the vector of rates. This algorithm has been implemented in an open-source Python package \verb|cgkmc|, which is available on the Python Package Index.

It is important to note that dynamically updating $\nu_t$ in the fashion above is purely a trick to prevent the crystal from growing too large. This has potentially unfavorable consequences, e.g. the final crystal morphology is independent of $D$ and $n_\infty$, which only affect the timescale of the simulation. However, it is ultimately necessary to inhibit the crystal's ability to grow to an equilibrium shape. This limitation has been elaborated upon in Sec. \ref{sec:limits}.
\section{Simulation Parameters}

The crystal structure of bulk PETN and corresponding nearest-neighbor interaction energies were obtained from a previous study by Singh. et al.\cite{doi:10.1021/acs.cgd.3c01487}. In this study, bulk PETN was created from the Cambridge Structural Database\cite{Groom2016-br} entry PERYTN12\cite{Zhurova2006-yc}, and lattice parameters were calculated by relaxing the supercell using density-functional tight binding (DFTB)\cite{10.1063/1.5044797} implemented within the LATTE \cite{LATTE}-LAMMPS\cite{THOMPSON2022108171} interface using the \textit{lanl}31+D DFTB parametrization\cite{10.1063/1.5063385}\cite{doi:10.1021/acs.jpclett.2c02701} which was developed for organic molecules and applied to the modeling of energetic materials~\cite{10.1063/1.5063385,Perriot2020,doi:10.1021/acs.jpclett.2c02701,Manner2024}. Additionally, 25 surface energies were calculated from DFTB, and fit as a function of the number of nearest-neighbors lost per unit surface area when creating a surface, yielding nearest-neighbor interaction energies; it was found that surface energies were reproduced by taking into account the first and second nearest neighbor interactions only. These structures and interaction parameters are summarized in Table \ref{tab:petn-params}.

\begin{table}[h]
    \centering
    \begin{tabular}{ll}
    \hline
    \textbf{Quantity} & \textbf{Value} \\
    \hline
    Unit cell structure & tetragonal \\
    Lattice constants & $a = b = \SI{9.087}{\AA}$\\
                      & $c = \SI{6.738}{\AA}$ \\
    Basis & $\{(0, 0, 0), (1/2, 1/2, 1/2)\}$ \\
    Space group & $P\overline{4}2_1c$ \\
    \hline
    Interaction energies & $\varepsilon_1 = \SI{-0.294}{eV}$ \\
                         & $\varepsilon_2 = \SI{-0.184}{eV}$ \\ 
                         & $\varepsilon_3 = \SI{-0.002}{eV}$ \\
    \hline
    Neighbor cutoffs & $\delta_1 = \SI{7.0}{\AA}$ \\
                     & $\delta_2 = \SI{7.5}{\AA}$ \\
                     & $\delta_3 = \SI{9.5}{\AA}$ \\
    \hline
    \end{tabular}
    \caption{Structure and interaction parameters for a PETN lattice.\cite{doi:10.1021/acs.cgd.3c01487}.}
    \label{tab:petn-params}
\end{table}

Note that the third-nearest neighbor interaction parameter $\varepsilon_3$ is much smaller than $\varepsilon_2$ in magnitude, meaning that third-nearest neighbor interactions can be neglected. We have repeated this work with $\varepsilon_3 = 0$ to confirm this, reproducing indistinguishable results. However, to remain consistent with Singh et al.\cite{doi:10.1021/acs.cgd.3c01487}, we use all three interaction parameters.

We then initialize a $62\times 62\times 140$ PETN lattice, denoted by the lattice positions $(\mathbf{r}_1, \cdots, \mathbf{r}_n)$. Then, we calculate the pairwise distances between each lattice point using the sliding-midpoint method \cite{DBLP:journals/corr/cs-CG-9901013, scipy} with a cutoff distance of $\SI{9.5}{\AA}$. Using these distances, we initialize the interaction matrix $\mathbf{Q}$ using Eq.~\eqref{eq:interaction-matrix} with cutoffs $(\delta_1, \delta_2, \delta_3)$ and interaction energies $(\varepsilon_1, \varepsilon_2, \varepsilon_3)$. Then, we initialize a spherical seed with a diameter of $\SI{15}{nm}$, and time-evolve the system according to Alg. \ref{alg:kmc-simplified}, using parameters from Table \ref{tab:implicit-variables}.

\begin{table}[H]
    \centering
    \begin{tabular}{|c|c|c|}
    \hline
    quantity & variable & value \\
    \hline
    beta & $\beta$ & $\SI{38.68}{eV^{-1}}$ \\
    \hline
    diffusivity & $D$     & $\SI{1.0e11}{\AA^2/s}$\\
    \hline
    solubility limit & $n_\infty$     & $\SI{1e-4}{\AA^{-3}}$\\
    \hline
    density & $\rho$ & $\SI{3.594e-3}{\AA^{-3}}$\\
    \hline
    desired final size & $N$ & $40,000$ \\
    \hline
    initial radius & $R_\text{init}$ & $\SI{75}{\AA}$ \\
    \hline
    number of iterations & $I$ & $1,000,000$ \\
    \hline
    \end{tabular}
    \caption{Implicit parameters used in KMC routine. $\beta = \SI{38.68}{eV^{-1}}$ corresponds to $\SI{300}{K}$, and $\rho = 2/(abc)$.}
    \label{tab:implicit-variables}
\end{table}

Both parameters $D$ and $n_\infty$ are functions of the solvent that the crystal is grown in. However, in our KMC routine, both of these parameters only change the timescale of the simulation, and not the steady state morphology of the crystal, which is elaborated upon in Sec. \ref{sec:results}. As such, we have selected these two parameters arbitrarily.
\section{Results}\label{sec:results}

\begin{figure}[H]
    \centering
    \includegraphics[width=\linewidth]{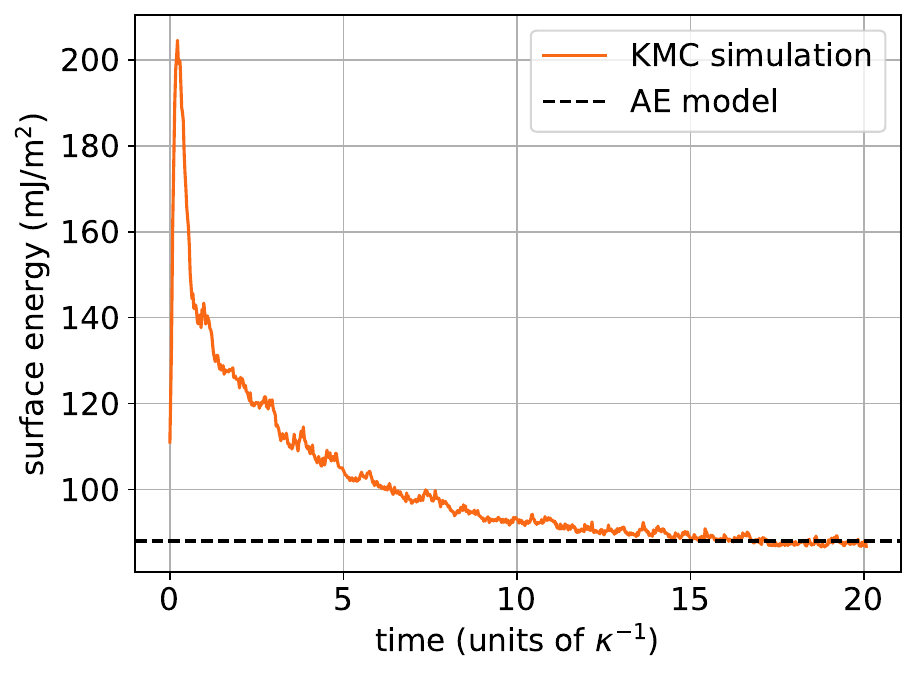}
    \caption{Surface energy as a function of time compared to AE model surface energy $\gamma_\text{AE}$.}
    \label{fig:convergence}
\end{figure}

The KMC routine outputs the time series $\mathbf{x}(t)$. From this time series, we calculate the total energy $E$ and surface area $S$ as functions of time. The surface area $S(t)$ is calculated via the $\alpha$-shape method\cite{Stukowski2014} in OVITO\cite{Stukowski_2010} with smoothing level 100. The surface energy $\gamma(t)$ is then:

\begin{equation}
    \gamma(t) = \frac{E(t) - E_\text{coh}N(t)}{S(t)}
\end{equation}

where $E_\text{coh}$ is the cohesive energy, which we calculated to be $\SI{-1.034}{eV}$ for the PETN parameters, $N(t) = \|\mathbf{x}(t)\|_1$ is the number of occupied sites at time $t$, and $\|\cdot\|_1$ denotes the $L^1$-norm. The resulting surface energy is benchmarked against the surface energy predicted by the aforementioned AE model (Fig. \ref{fig:convergence}):

\begin{equation}
    \gamma_\text{AE} = \sum_{hk\ell}\phi_{hk\ell}\gamma_{hk\ell}
\end{equation}

where $\phi_{hk\ell}$ is the area fraction of the $(hk\ell)$ surface computed by the AE model, and $\gamma_{hk\ell}$ is the surface energy. Both of these sets of parameters were computed by Singh et al.\cite{doi:10.1021/acs.cgd.3c01487}, yielding $\gamma_\text{AE} \approx \SI{88}{mJ/m^2} \approx \SI{5.5}{meV/\AA^2}$.

\begin{figure}[H]
    \centering
    \includegraphics[width=\linewidth]{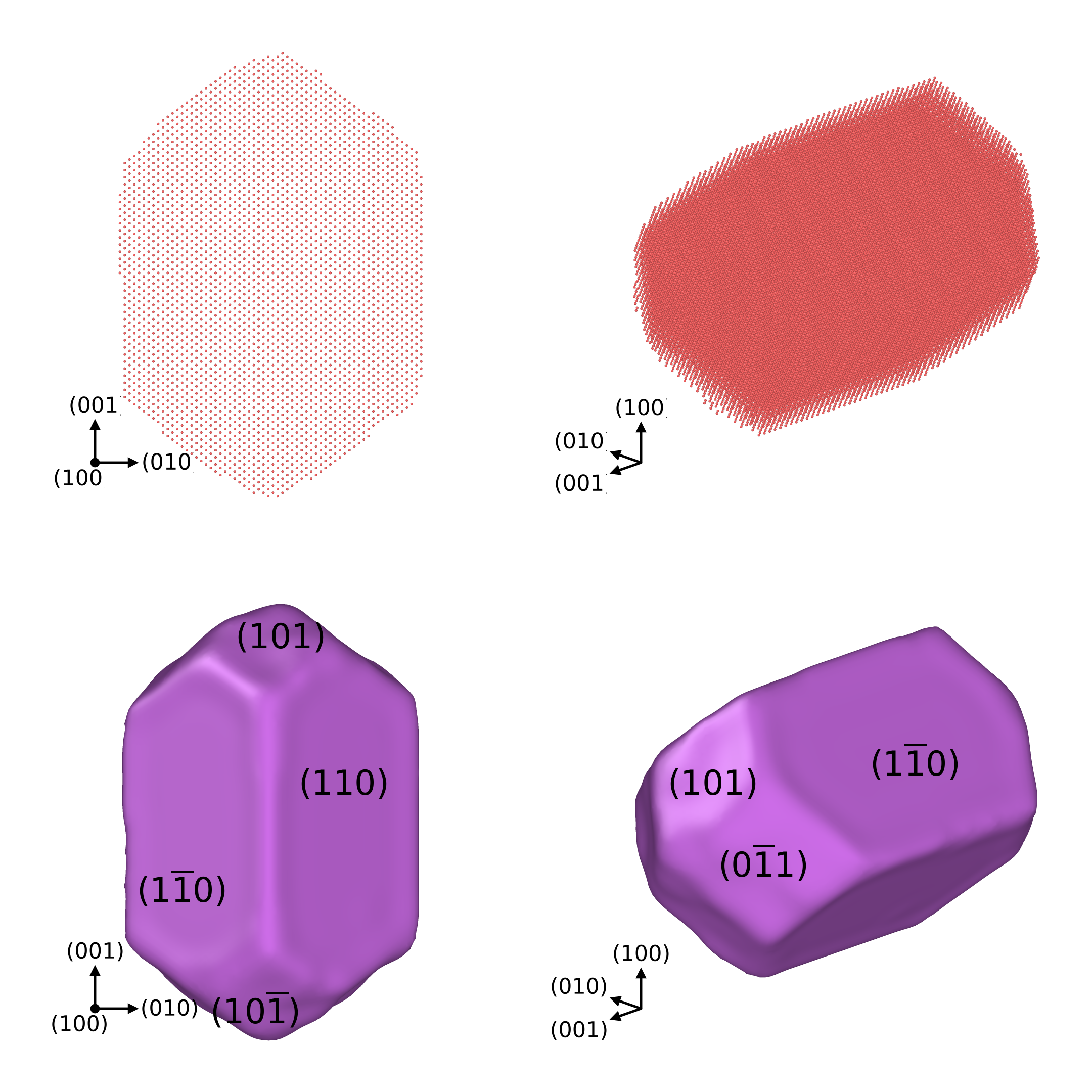}
    \caption{Final crystal shape with two different orientations. Bottom is with surface mesh built with the $\alpha$-shape method in OVITO with radius $\SI{7}{\AA}$ and smoothing level $100$.}
    \label{fig:shape}
\end{figure}

Our KMC routine predicts a total surface energy that asymptotically agrees with the AE model (Fig. \ref{fig:convergence}), while additionally providing a dynamical pathway to the steady-state configuration. Similarly, the emergent planes are in agreement: the AE model predicts dominant $\{110\}$ and $\{101\}$ planes, in exact agreement with our KMC routine (Fig. \ref{fig:shape}).
\section{Discussion}\label{sec:limits}

Although our KMC results for PETN agree well with the AE model, as well as experimental observations\cite{doi:10.1021/acs.cgd.3c01487}, there are various limitations of our proposed routine.

Firstly, our routine for dynamically updating $\nu_t$ (Eq.~\eqref{eq:evaporation-prefactor}) has unphysical consequences. Namely, the steady-state morphology is independent of both $D$ and $n_\infty$, since both the evaporation and adsorption events are proportional to $\kappa\propto Dn_\infty$. This has the consequence that the morphology is independent of the solvent type. This is in stark disagreement with experiments, which show that the aspect ratio of PETN is strongly dependent on the solubility limit of PETN in the solvent\cite{Davis2025}. A potential fix would be to replace $\nu_t$ with a constant rate prefactor $\nu$. However, we found that a constant prefactor yields a thin rod-like structure with only $\{110\}$ surfaces, with minimal surface energy $\gamma = \gamma_{110} < \gamma_\text{AE}$, in disagreement with experiment. This morphology is correct in the surface energy-minimization sense, but is clearly incorrect in the growth-kinetics sense. We suspect that a carefully parametrized, non-constant rate prefactor could correctly predict growth kinetics, but this parametrization is likely non-trivial, and is thus saved for a future work. However, in principle, our routine is very amenable to this extension, since changing the prefactor only affects the calculation of the rates $\mathbf{R}$.

Secondly, our routine explicitly excludes surface diffusion and solvation effects. We expect both effects could play an important role in predicting more complex morphologies. However, local features of the surface, such as roughness of the surface, can have strong solvation effects on the surface\cite{10.1063/1.475973, doi:10.1021/acs.jpcc.5b04598} and surface diffusion\cite{PhysRevB.56.12135, 10.1063/1.1475774, PhysRevB.81.195421}. These local features are relatively expensive to calculate on-the-fly, significantly slowing down a KMC simulation. If these effects are properly parametrized, though, our routine largely stays the same, except with additional energy barriers for solvation and surface diffusion candidate events.

Thirdly, our routine predicts the growth of a single crystal, and can therefore not include crowding effects out-of-the-box. Although one could instead initialize multiple seeds, rather than one spherical seed, these seeds would immediately fuse into singular crystals once their interfaces come in contact with one another. This could potentially be remedied by the inclusion of solvation layers and/or running simulations on super-lattices encoding multiple orientations, but this is very non-trivial, and is similarly saved for a future work.

Lastly, our energetics model assumes that the solid phase can be modeled with only pairwise intermolecular interactions. This model is especially useful in the context of our routine, since a local intermolecular energy naturally arises from it. For more complex systems, a simple quadratic form might not be sufficient, and more sophisticated routines might need to be employed to encode locality, e.g. using moment tensors for describing local environments\cite{10.1063/1.5005095}.

\section{Conclusions}

In this work, we developed a graph-based KMC simulation routine to predict crystal morphology, depending only on intermolecular interaction energies within the solid phase. This routine was then implemented within the open-source \verb|cgkmc| package. We then use \verb|cgkmc| to predict the morphology of a PETN crystal, in excellent agreement with the AE model, while also providing information about the intermediate growth dynamics. We additionally sketch the potential future inclusion of more complex physics - namely surface diffusion, solvent effects, surface roughness, and complex intermolecular interactions.
\section{Data Availability}

\verb|cgkmc|'s source code is available on GitHub\cite{github_repo}, installable from PyPI\cite{cgkmc}, and fully documented through GitHub Pages\cite{cgkmc_docs}. The PETN simulation ran by \verb|cgkmc| for this study is included in the library's documentation, and the raw simulation data is available upon request.
\section{Acknowledgements}

This work was supported by the Laboratory Directed Research and Development program of Los Alamos National Laboratory under project no. 20220431ER. This research used resources provided by the Los Alamos National Laboratory Institutional Computing Program. Los Alamos National Laboratory is operated by Triad National Security, LLC, for the National Nuclear Security Administration of U.S. Department of Energy (contract no. 89233218CNA000001).

Additionally, this material is based on work supported by the National Science Foundation under Grant Nos. MRI\# 2024205, MRI\# 1725573, and CRI\# 2010270 for allotment of compute time on the Clemson University Palmetto Cluster.
\section{Disclaimer}

Any opinions, findings, and conclusions or recommendations expressed in this material are those of the author(s) and do not necessarily reflect the views of the National Science Foundation.
\appendix

\section{Appendix}

\subsection{Adsorption Rate}\label{sec:ads-rate}

For a solution occupying region $\mathcal{S}$ and a solid crystal occupying region $\Omega\subseteq\mathcal{S}$, we can calculate the flux on the surface of the crystal by modeling using the steady-state diffusion equation:

\begin{equation}
    \nabla^2 n = 0 \text{ st. } n\big|_{\partial\Omega} = 0 \text{ and } n\big|_{\partial\mathcal{S}} = n_\infty
\end{equation}

Here, $\partial\Omega$ represents the surface of the crystal, and $\partial\mathcal{S}$ represents the extrema of the solution, i.e. very far away from the crystal, and $n$ is the spatial concentration profile of solute molecules in the solution. Then, $n_\infty$ is the far-field concentration. In spherical coordinates, assuming that the crystal is spherical with radius $R$:

\begin{equation}
    \nabla^2 n = 0 \text{ st. }
    \begin{matrix}
        n(R, \theta, \varphi) = 0
        \\ n(\infty, \theta, \varphi) = n_\infty
    \end{matrix}
\end{equation}

Additionally, assuming that the liquid phase is isotropic, there will be no $\theta$ and $\varphi$ dependencies. So, separating variables $n(r, \theta, \varphi) = n_1(r)n_2(\theta)n_3(\varphi)$, where $r$ is the distance from the center of the crystal, yields:

\begin{equation}
    \frac{\partial}{\partial r}\left(r^2\frac{\partial n_1}{\partial r}\right) = 0 \text{ st. } n_1(R) = 0 \text{ and } n_1(\infty) = n_\infty
\end{equation}

which yields:

\begin{equation}
    n_1(r) = \frac{c_1}{r} + c_2
\end{equation}

where $c_1$ and $c_2$ are integration constants. Plugging in boundary conditions yields:

\begin{equation}
    n_1(r) = n_\infty\left(1-\frac{R}{r}\right)
\end{equation}

which has surface flux:

\begin{equation}
    J(R) = -Dn_1'(R) = -\frac{Dn_\infty}{R}
\end{equation}

The number of molecules going to the each lattice site on the surface is then:

\begin{equation}
    \dot n_\text{adsorption} = -\frac{J\cdot 4\pi R^2}{\sigma\cdot 4\pi R^2} = -\frac{J}{\sigma} = \frac{Dn_\infty}{\sigma R}
\end{equation}

where $\sigma$ is the surface density of sites, such that $\sigma\cdot 4\pi R^2$ is the number of sites on the surface.

\subsection{Evaporation barrier}\label{sec:evap-barrier}

The energy difference between states $\mathbf{y}$ and $\mathbf{x}$ is:

\begin{equation}
    \begin{aligned}
        \Delta E &= \frac{1}{2}\mathbf{y}^\intercal\mathbf{Q}\mathbf{y} - \frac{1}{2}\mathbf{x}^\intercal\mathbf{Q}\mathbf{x} \\
        &= \frac{1}{2}\sum_{ij}Q_{ij}(y_i y_j - x_i x_j)
    \end{aligned}
\end{equation}

Suppose we want to either evaporate or adsorb a molecule at site $k$. Then, the new state vector $\mathbf{y}$ is:

\begin{equation}
    y_i = \begin{cases}
        1 - x_i & i = k \\
        x_i & i\neq k
    \end{cases}
\end{equation}

So, $y_iy_j \neq x_ix_j$ only if $i = k$ or $j = k$. Therefore, assuming that $Q_{ii} = 0$:

\begin{equation}
    \begin{aligned}
        \Delta E &= \frac{1}{2}\sum_{i\neq k} Q_{ik}(y_i y_k - x_i x_k) \\
        &+\frac{1}{2}\sum_{j\neq k} Q_{kj}(y_k y_j - x_k x_j)
    \end{aligned}
\end{equation}

The $i = k$ terms and $j = k$ terms are excluded because $Q_{kk} = 0$. Additionally, since $\mathbf{Q}$ is symmetric:

\begin{equation}
    \begin{aligned}
        \Delta E &= \sum_{i\neq k} Q_{ik}(y_i y_k - x_i x_k)\\
        &=\sum_{i\neq k} Q_{ik}x_i(y_k - x_k) \\
        &=(1 - 2x_k)\mathbf{Q}_k\cdot \mathbf{x}
    \end{aligned}
\end{equation}

If site $k$ is being evaporated, then $\mathbf{x}_k = 1$, meaning $\Delta E = -\mathbf{Q}_k\cdot\mathbf{x}$, which completes the proof for the evaporation barrier. Similarly, the change in energy of adsorption at site $k$ is $\Delta E = \mathbf{Q}_k\cdot\mathbf{x}$.

\bibliography{bibfile.bib}

\end{document}